\documentclass[12pt, a4paper]{article}
\usepackage[margin=25truemm]{geometry}
\usepackage{graphicx, amsmath, amssymb, comment, pgfplots, cite}
\usepackage[colorlinks=true,citecolor=blue,linkcolor=blue,urlcolor=blue]{hyperref}

\pgfplotsset{compat=1.18}

\begin{document}

\begin{titlepage}

\begin{center}
  \begin{flushright}

\end{flushright}

\vskip 2cm

\begin{center}
{\bf\Large Gamma-Ray Signatures of\\ Thermal Misalignment Dark Matter} \\

\vskip 1.5cm

\renewcommand*{\thefootnote}{\fnsymbol{footnote}}
Koichi Hamaguchi$^{a,b}$\footnote{hama@hep-th.phys.s.u-tokyo.ac.jp},
Ryoichiro Hayakawa$^{a}$\footnote{rhayakawa@hep-th.phys.s.u-tokyo.ac.jp}
and
Hiroki Takahashi$^{a}$\footnote{takahashi@hep-th.phys.s.u-tokyo.ac.jp}
\end{center}

\vskip 0.8cm

{\it $^a$Department of Physics, University of Tokyo, Bunkyo-ku, Tokyo 113--0033, Japan} \\[2pt]
{\it ${}^b$Kavli Institute for the Physics and Mathematics of the Universe (Kavli IPMU), University of Tokyo, Kashiwa 277--8583, Japan} 

\date{\today}

\vskip 1.5cm

\begin{abstract}
Thermal misalignment is a viable dark matter scenario where the misalignment of a dark matter scalar, feebly coupled to the Standard Model particles, is generated through thermal effects from the primordial plasma.
In this framework, the scalar is generically metastable, and its decay can leave observable signatures.
In this work, we focus on the case in which the scalar $\phi$ is coupled to photons through $\phi F^{\mu\nu} F_{\mu\nu}$, and examine its observational signatures.
We find that current gamma-ray constraints place a robust upper bound on the scalar mass of  $\mathcal O(1)\,\mathrm{GeV}$. 
We also find that future observations can further probe the parameter region, particularly in the MeV--GeV range, an energy band expected to be explored by various gamma-ray observatories in the coming decades.
\end{abstract}

\end{center}
\end{titlepage}

\renewcommand*{\thefootnote}{\arabic{footnote}}
\setcounter{footnote}{0}

\section{Introduction}
Although there is mounting evidence for the existence of dark matter (DM)~\cite{ParticleDataGroup:2024cfk}, its fundamental nature remains unknown. From the viewpoint of minimality, one of the simplest possibilities is to extend the Standard Model (SM) by adding a single real scalar field, $\phi$, as a DM candidate. 
At the renormalizable level, the only possible interaction between $\phi$ and the SM is the Higgs portal coupling, and this minimal scenario has been studied extensively.
However, its simplest thermal relic realization~\cite{Silveira:1985rk,McDonald:1993ex,Burgess:2000yq} is now severely constrained by direct detection experiments~\cite{EscuderoAbenza:2025cfj}.

An alternative possibility is that $\phi$ couples to the SM through a dimension-five operator of the form $\dfrac{\phi}{M} {\cal O}_{\rm SM}
$, where ${\cal O}_{\rm SM}$ is a dimension-four operator composed of SM fields and $M$ is a large mass scale characterizing the effective interaction.
In the early Universe, even if $\phi$ itself is not thermalized, interactions of this kind generally lead to finite-temperature corrections to the effective potential of $\phi$ when SM fields entering ${\cal O}_{\rm SM}$ are in the thermal bath.
In particular, thermal effects can induce linear terms in $\phi$, shifting the potential minimum and thereby significantly affecting the cosmological evolution of $\phi$, as discussed in early studies~\cite{Buchmuller:2003is,Buchmuller:2004xr,Buchmuller:2004tz}. 
More recently, this effect has been applied to dark matter production, leading to the thermal misalignment DM scenario~\cite{Batell:2021ofv,Chun:2021uwr}.
The implications of such scenarios have been discussed in a variety of contexts~\cite{Batell:2022qvr,Alachkar:2024crj,Cyncynates:2024bxw,Cyncynates:2024ufu,Batell:2026sml}.

As long as the coupling to the SM is linear in $\phi$, thermal misalignment DM is generically metastable and therefore behaves as decaying DM.
In this paper, we focus on the case where the scalar is coupled to the photon via the operator $\phi F^{\mu\nu} F_{\mu\nu}$, and study its observational signatures. 
The scalar can then decay at tree level into two photons, leading to potentially observable photon signals. We show that current gamma-ray constraints already place a robust upper bound of $\mathcal O(1)~\mathrm{GeV}$ on the scalar mass.
This makes the MeV--GeV range particularly important, as it is expected to be probed by various observations in the coming decades, such as 
COSI~\cite{Tomsick:2023aue}, GECCO~\cite{Orlando:2021get},
e-ASTROGAM~\cite{e-ASTROGAM:2017pxr},
AMEGO~\cite{Kierans:2020otl},
AMEGO-X~\cite{Caputo:2022xpx},
MAST~\cite{Dzhatdoev:2019kay},
AdEPT~\cite{Hunter:2013wla},
PANGU~\cite{Wu:2014tya},
and
GRAMS~\cite{GRAMS:2021tax}.
We show that these future observations can further probe the parameter region predicted in the thermal misalignment DM scenario.

The rest of this paper is organized as follows. In Section~\ref{sec:thermal misalignment}, we introduce the setup and briefly explain the thermal misalignment production of dark matter.
Section~\ref{sec:gamma-ray} discusses the gamma-ray signals from the decay of the DM scalar, presenting a robust upper bound on the mass as well as prospects for future observations.
Section~\ref{sec:summary} is devoted to a summary and discussion.


\section{Thermal Misalignment Dark Matter Coupled to Photons}\label{sec:thermal misalignment}

In this section, the setup of thermal misalignment DM coupled to photons is introduced, and the basic mechanism relevant for the present work is briefly reviewed. 
The discussion below builds on the general analyses of Ref.~\cite{Cyncynates:2024bxw}, with some generalizations and modifications appropriate for the present work.

As a minimal extension to the SM, we consider the following Lagrangian for a CP-even real scalar field $\phi$ with a linear coupling to the electromagnetic field strength $F_{\mu\nu}$,
\begin{align}
\mathcal{L}_{\phi} 
&=\frac{1}{2}\partial_\mu \phi \partial^\mu \phi- \frac{1}{2}m^2 \phi^2-\frac{\phi}{M} F^{\mu\nu}F_{\mu\nu},\label{eq:model}
\end{align}
where $m$ denotes the scalar mass, $M$ is a scale parameterizing the dimension-five operator, and $F_{\mu\nu}$ is canonically normalized: its kinetic term reads $-\dfrac{1}{4}F^{\mu\nu}F_{\mu\nu}$.
As we will see, in the parameter region of interest, $M$ turns out to be larger than the Planck scale,
$M_{\rm P}\simeq 2.4\times10^{18}\,{\rm GeV}$.

Since the thermal misalignment mechanism relevant for the present work mainly operates at temperatures above the electroweak scale, it is convenient to embed the photon coupling in Eq.~\eqref{eq:model} into an electroweak-invariant form:
\begin{align}
\mathcal{L}_{\phi,{\rm EW}} 
&=\frac{1}{2}\partial_\mu \phi \partial^\mu \phi
- \frac{1}{2}m^2 \phi^2
-\frac{\xi}{\cos^2\theta_W}\frac{\phi}{M}B^{\mu\nu}B_{\mu\nu}
-\frac{1-\xi}{\sin^2\theta_W}\frac{\phi}{M}W^{a\mu\nu}W^a_{\mu\nu}
,\label{eq:model_EW}
\end{align}
where $B^{\mu\nu}$ and $W^{a\mu\nu}$ are the field strengths of the $U(1)_Y$ and $SU(2)_L$ gauge fields, respectively, $\theta_W$ is the weak mixing angle, and $\xi$ is a real parameter. 
Treating $\phi$ as a background field,\footnote{As noted above, $M$ is larger than the Planck scale, and thus the scalar $\phi$ does not enter the thermal bath.} these couplings make the electroweak gauge couplings $\phi$-dependent after canonical normalization of the gauge fields. To leading order in $\dfrac{\phi}{M}$, this amounts to
\begin{align}
g_Y^2(\phi) \simeq g_Y^2\left(1-\frac{4\xi}{\cos^2\theta_W}\frac{\phi}{M}\right),
\quad
g_2^2(\phi) \simeq g_2^2\left(1-\frac{4(1-\xi)}{\sin^2\theta_W}\frac{\phi}{M}\right),
\label{eq:effective_coupling_EW}
\end{align}
where $g_Y = \dfrac{e}{\cos\theta_W}$ and $g_2=\dfrac{e}{\sin\theta_W}$. 

In the early universe, where SM particles are in thermal equilibrium, the background scalar field acquires a thermal correction to its effective potential through the free energy of the plasma. 
The leading-order correction to the free energy of $SU(2)_L\times U(1)_Y$ is given by~\cite{Cyncynates:2024bxw,Kapusta:2006pm,Gynther:2005dj,Gynther:2006wq}
\begin{align}
    \mathcal{F}_2
    &= \left( 
    \frac{55}{576} 
    g_Y^2(\phi) 
    + \frac{43}{192} 
    g_2^2(\phi) 
    \right) T^4,
    \label{eq:F2}
\end{align}
for $T\gg 100~{\rm GeV}$.
This gives rise to the following effective potential $V_{\rm eff}$ for $\phi$ through Eq.~\eqref{eq:effective_coupling_EW}:\footnote{In Ref.~\cite{Cyncynates:2024bxw}, the coefficient of the thermal linear term was presented without making the dependence on the electroweak embedding parameter $\xi$ explicit. As will be seen below, however, an $\mathcal O(1)$ variation of $\xi$ can lead to a change in the DM abundance by more than two orders of magnitude. We therefore keep the $\xi$ dependence explicit in the following analysis.}
\begin{align}
    V_{\rm eff}(\phi) 
    &= \frac{1}{2}m^2\phi^2 + V_T(\phi), \label{eq:eff}\\
    V_T(\phi) 
    &\simeq  
    -\left(\frac{55}{144}\frac{\xi}{\cos^2\theta_W}g_Y^2
    + \frac{43}{48}\frac{1-\xi}{\sin^2\theta_W}g_2^2
    \right)\frac{\phi}{M}T^4,\label{eq:tp}
\end{align}
where we omitted the $\phi$-independent constant term.

The equation of motion for $\phi$ is given by
\begin{align}
\ddot{\phi} + 3H\dot{\phi} + \frac{\partial V_{\rm eff}(\phi)}{\partial \phi}=0,
\end{align}
where the dot denotes the derivative with respect to cosmic time $t$. For convenience, we define dimensionless variables 
\begin{align}
    x\equiv mt,\quad 
    y\equiv \frac{\phi}{\phi_*},\quad
    \phi_*\equiv \frac{\alpha M_{\rm P}^2}{M},\quad 
    \alpha\equiv \dfrac{e^2}{4\pi},
\end{align}
with which the equation of motion reads
\begin{align}
&y''+\frac{3}{2x}y'+y-\frac{Q(\xi)}{x^2} =0,\quad
Q(\xi) \equiv \left(0.64\xi+16.93(1-\xi)\right)
\frac{90}{\pi g_*},
\label{eq:Qdef}
\end{align}
where we have used $\sin^2\theta_W\simeq0.23$ in evaluating $Q(\xi)$.
We have also assumed a radiation dominated universe; the Hubble parameter $H$ is given by $H=\dfrac{1}{2t}$ and satisfies $3M_{\rm P}^2 H^2 = \dfrac{\pi^2}{30}g_*T^4$, where $g_*=106.75$ is the effective number of relativistic degrees of freedom.

\begin{figure}[t]
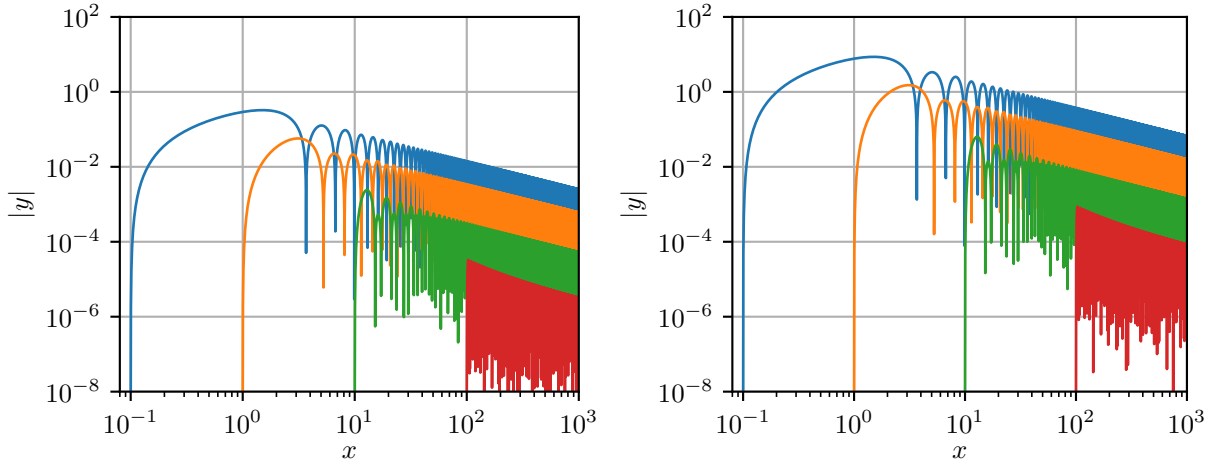

\centering
\begin{minipage}[h]{0.49\linewidth}
    \centering
    \input{y_time_evol_xi_1.pgf}
\end{minipage}
\begin{minipage}[h]{0.49\linewidth}
    \centering
    \input{y_time_evol_xi_0.pgf}
\end{minipage}
\caption{The evolution of $|y|$ for $x_i = 10^{-1},\ 1,\ 10,\text{and }10^2$ (blue, orange, green and red, respectively), $\xi=1\ \text{(left) and }0\ \text{(right)}$, and $y(x_i)=0$.
}
\label{fig:x_to_y}
\end{figure}

In Fig.~\ref{fig:x_to_y}, we show the evolution of $|y|$ 
for $\phi_{\rm ini}=0$ and $x_{\rm ini}=m t_{\rm ini}=10^{-1},\ 1,\ 10,$ and $10^2$ in cases $\xi=1\text{ and }0$.\footnote{For $x_{\rm ini}\lesssim 1$, the subsequent evolution is largely insensitive to the initial value $\phi_{\rm ini}$ as long as $\phi_{\rm ini}\lesssim \phi_*$. For $x_{\rm ini}\gtrsim 1$, the Hubble parameter during inflation is larger than the scalar mass, $H_{\rm inf} > m$, under the assumption of instantaneous reheating, and hence the scalar field $\phi$ is driven to the origin during inflation.}
Here $t_{\rm ini}$ denotes the cosmic time at reheating.
For simplicity, we assume instantaneous reheating in the following discussion.

As can be seen in the figure, the cosmological evolution of the scalar is distinct for the cases of $x_{\rm ini} \gtrsim 1$ and $x_{\rm ini} \lesssim 1$.
For $x_{\rm ini} \gtrsim 1$, i.e. $t_{\rm ini} \gtrsim m^{-1}$, the scalar begins to oscillate almost immediately after reheating in response to the shift of the potential minimum.
On the other hand, for $x_{\rm ini} \lesssim 1$, i.e. $t_{\rm ini} \lesssim m^{-1}$, the scalar field continues to evolve until $t \sim m^{-1}$ toward $\phi_*$, which is a characteristic field scale near the minimum of the effective potential $V_{\rm eff}(\phi)$ at $x\simeq 1$. 

In either case, the thermal potential becomes negligible at late times, and the oscillation of $\phi$ inevitably generates a relic scalar abundance, which may explain the observed DM density.
In Ref.~\cite{Cyncynates:2024bxw}, a detailed study of such a scenario is presented, which shows that the scalar produced by thermal misalignment can be a viable DM candidate in a wide parameter region with masses $\mathcal{O}(10^{-11})~{\rm eV} \lesssim m \lesssim \mathcal{O}(10^{14})~{\rm eV}$.
In the next section, we focus on the mass region $\mathcal{O}(10^{-2})~{\rm MeV} \lesssim m \lesssim \mathcal{O}(10^{1})~{\rm GeV}$, and discuss the testability of the model based on gamma-ray signals.

\section{Gamma-Ray Signatures of Thermal Misalignment Dark Matter}\label{sec:gamma-ray}

Since the scalar field is coupled to photons in our setup, it can decay into photon pairs. 
The rate of such decay must be sufficiently suppressed for the scalar field to survive until the present day. 
Moreover, even if the scalar is long-lived to be the DM, it can still emit gamma-ray signals, which can be a smoking gun signature for the scenario.
In this section, we compare the predictions of this scenario with current constraints and discuss its prospects for future detection.\footnote{Prospective sensitivities for smaller mass regions are discussed in Ref.~\cite{Cyncynates:2024bxw}.}

\begin{figure}[t] 
    \centering
    \input{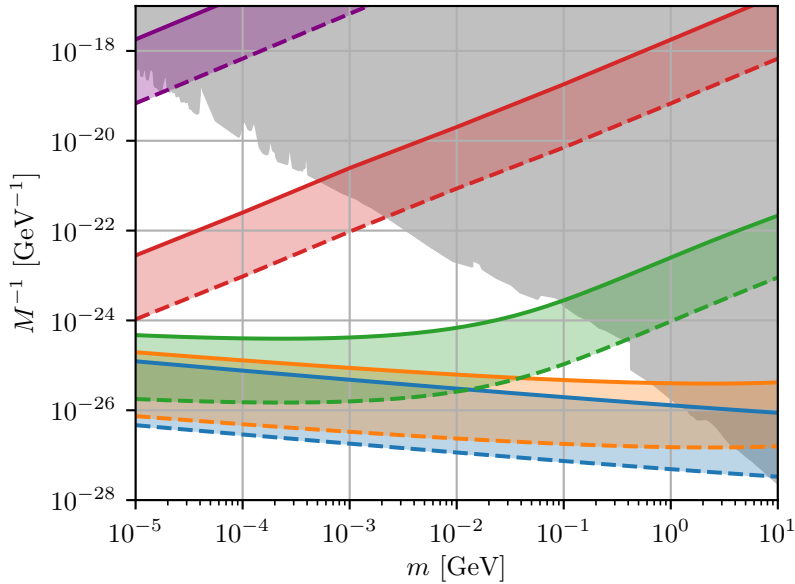}
    \caption{The parameter space for the scalar $\phi$ coupled to photons in the $(m,M)$-plane.
    The colored lines depict the values of the parameters that explain the observed DM abundance for $T_R=10^{12}\ \text{GeV}$ (blue), $10^{10}\ \text{GeV}$ (orange), $10^{8}\ \text{GeV}$ (green), $10^{6}\ \text{GeV}$ (red) and $10^{4}\ \text{GeV}$ (purple). 
    For each reheating temperature, the solid and dashed lines correspond to $\xi=1$ and $\xi=0$, respectively, and the region between them indicates the intermediate values of $\xi$.
    The gray shaded regions show current constraints from
Fermi-LAT~\cite{Fermi-LAT:2015kyq}, INTEGRAL/SPI~\cite{Fischer:2022pse} with NFW DM profile,  COMPTEL/EGRET~\cite{Essig:2013goa},
and NuSTAR~\cite{Roach:2022lgo}.}
    \label{fig:m-Mplane}
\end{figure}

\begin{figure}[t] 
    \centering
    \input{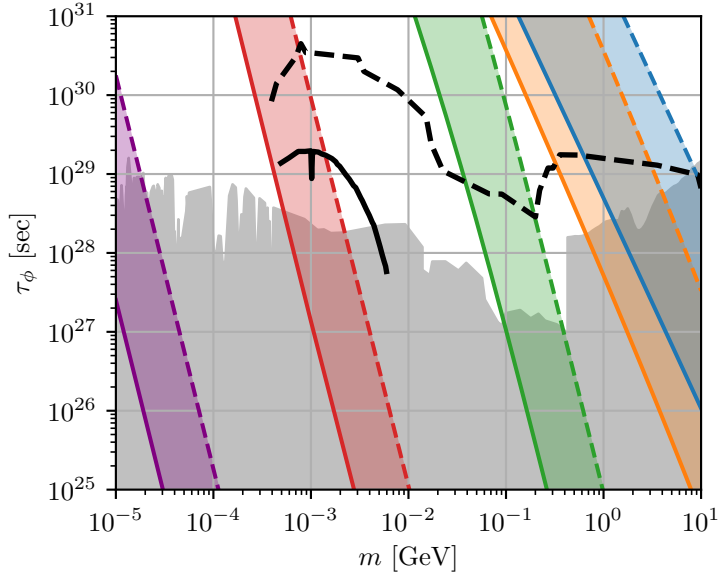}
    \caption{The predicted lifetime $\tau_\phi$ of thermal misalignment DM $\phi$ as a function of its mass $m$, shown in the $(m, \tau_\phi)$-plane.
    The colored lines correspond to the same $(T_R,\xi)$ pairs as in Fig.~\ref{fig:m-Mplane}.
    The gray shaded region indicates the same current constraints as in Fig.~\ref{fig:m-Mplane}.
    The thick black lines show the projected sensitivities of future gamma-ray observations: 
    COSI~\cite{Caputo:2022dkz} (solid) and other proposed missions summarized in Ref.~\cite{ODonnell:2024aaw} (dashed). 
    }
    \label{fig:m-lifetimeplane}
\end{figure}

The tree-level decay rate of $\phi$ into a pair of photons is given by~\footnote{Our result does not agree with the expression shown in Ref.~\cite{Cyncynates:2024bxw}.\\ Due to the relation $M^{-1}=\dfrac{d_{\alpha_{\rm EM}}^{(1)}}{4\sqrt{2}M_{\rm P}}=\dfrac{d_e^{(1)}}{2\sqrt{2}M_{\rm P}}$, the decay rate can be written in terms of $d^{(1)}_e$ as 
\begin{align}
\Gamma_{\phi\to \gamma\gamma} =\frac{1}{4} \frac{(2d_e^{(1)})^2 m^3}{32\pi M_p^2},
\end{align}
which differs from the one shown in Ref.~\cite{Cyncynates:2024bxw} by a factor of $4$.}
\begin{align}
    \Gamma(\phi\to\gamma\gamma)=\frac{m^3}{4\pi M^2}.\label{eq:decayrate}
\end{align}
In Fig.~\ref{fig:m-Mplane}, we show the parameter space for the scalar $\phi$ coupled to photons in the $(m,M)$-plane, focusing on 
the mass range $\mathcal{O}(10^{-2})~{\rm MeV} \lesssim m \lesssim \mathcal{O}(10^{1})~{\rm GeV}$.
The colored lines depict the values of the parameters that explain the observed DM abundance for given $(T_R, \xi)$.
For illustration, we focus on the representative range $0 \le \xi \le 1$: for each $T_R$
\footnote{The values of $T_R$ are chosen so that the contours for correct relic abundance of DM lie in the range of $m$ of interest, $10^{-5}\ \mathrm{GeV}<m<10\ \mathrm{GeV}$, in Fig.~\ref{fig:m-lifetimeplane}. For $T_R \gg 10^{10}\ \mathrm{GeV}$, the DM abundance depends on $T_R$ only logarithmically.
}, the solid and dashed lines correspond to $\xi=1$ and $\xi=0$, respectively, and the shaded region between them indicates intermediate values of $\xi$. However, in general, $\xi$ is not restricted to this interval. If $Q(\xi)$ approaches 0 (i.e.\ $\xi \simeq 1.04$, see Eq.~\eqref{eq:Qdef}), $M^{-1}$ needs to take a larger value; if $|\xi| \gg 1$, $M^{-1}$ must be smaller. 
The gray shaded regions show current constraints from
Fermi-LAT~\cite{Fermi-LAT:2015kyq}, INTEGRAL/SPI~\cite{Fischer:2022pse} with NFW DM profile,  COMPTEL/EGRET~\cite{Essig:2013goa},
and NuSTAR~\cite{Roach:2022lgo}.
We highlight that the current experimental bounds place a robust upper limit on the DM mass: $m \lesssim \mathcal{O}(1)$ GeV\footnote{Ref.~\cite{Cyncynates:2024bxw} noted that decays into photons can be constrained by astrophysical observations, but the relevant gamma-ray bounds, such as those from Fermi-LAT~\cite{Fermi-LAT:2015kyq}, were not incorporated into the parameter-space analysis. In the present work, we include these bounds explicitly, and they are crucial for deriving the upper bound on the scalar mass.
}. For example, $m\leq8.3\ \mathrm{GeV}$ in the case of $(T_R,\xi)=(10^{12}\ \mathrm{GeV},0)$.

For future prospects, Fig.~\ref{fig:m-lifetimeplane} shows the predicted lifetime of thermal misalignment DM as a function of its mass, together with current constraints and projected sensitivities of various gamma-ray searches.
The thick black lines show the projected sensitivities of future gamma-ray observations: 
    COSI~\cite{Caputo:2022dkz} (solid) and other proposed missions summarized in Ref.~\cite{ODonnell:2024aaw} (dashed).
As can be seen from the figure, future gamma-ray observations can further probe the parameter region predicted in the thermal misalignment DM scenario. This provides further motivation for gamma-ray searches.

\section{Summary and Discussion}\label{sec:summary}
In this paper, we have studied gamma-ray signals coming from the decaying scalar dark matter coupled to photons through the dimension-five operator, produced by thermal misalignment. We first clarified that the abundance is determined not only by the reheating temperature $T_R$ but also by a parameter associated with the ambiguity in electroweak embedding $\xi$, whose role has not been discussed in previous literature. 
In particular, we showed that $\xi$ can affect the DM abundance by more than two orders of magnitude. 

We then investigated the observational consequences of the DM decay into photons. By reexamining the existing gamma-ray constraints, we found a robust upper bound on the scalar mass, $m \lesssim \mathcal{O}(1)$ GeV.
We also projected the prediction of the thermal misalignment DM onto the mass-lifetime plane and showed that future gamma-ray searches can further probe the parameter region in the MeV-GeV range. 
This provides further motivation for those proposed gamma-ray observations.

Gamma-ray signatures are expected to be produced in a number of similar, minimal setups other than the photon coupling, although the decay rates are loop-suppressed. We anticipate that some of the parameter spaces of those models can also be explored by future gamma-ray observatories considered in this work. 
It would also be important to investigate possible ultraviolet completions of the effective interactions considered here. 
A detailed study of these issues, as well as other signals in each setup, is left for future work.


\section*{Acknowledgements}

This work was supported by JSPS KAKENHI Grant Numbers 24H02244, 24K07041 (KH), and 24KJ0913 (HT).


\bibliographystyle{JHEP}
\bibliography{biblio}

@article{Buchmuller:2004tz,
    author = "Buchmuller, Wilfried and Hamaguchi, Koichi and Lebedev, Oleg and Ratz, Michael",
    title = "{Maximal temperature in flux compactifications}",
    eprint = "hep-th/0411109",
    archivePrefix = "arXiv",
    reportNumber = "DESY-04-216",
    doi = "10.1088/1475-7516/2005/01/004",
    journal = "JCAP",
    volume = "01",
    pages = "004",
    year = "2005"
}

@article{Buchmuller:2004xr,
    author = "Buchmuller, Wilfried and Hamaguchi, Koichi and Lebedev, Oleg and Ratz, Michael",
    title = "{Dilaton destabilization at high temperature}",
    eprint = "hep-th/0404168",
    archivePrefix = "arXiv",
    reportNumber = "DESY-04-062",
    doi = "10.1016/j.nuclphysb.2004.08.031",
    journal = "Nucl. Phys. B",
    volume = "699",
    pages = "292--308",
    year = "2004"
}

@book{Kapusta:2006pm,
    author = "Kapusta, J. I. and Gale, Charles",
    title = "{Finite-temperature field theory: Principles and applications}",
    doi = "10.1017/CBO9780511535130",
    isbn = "978-0-521-17322-3, 978-0-521-82082-0, 978-0-511-22280-1",
    publisher = "Cambridge University Press",
    series = "Cambridge Monographs on Mathematical Physics",
    year = "2011"
}

@article{Batell:2021ofv,
    author = "Batell, Brian and Ghalsasi, Akshay",
    title = "{Thermal misalignment of scalar dark matter}",
    eprint = "2109.04476",
    archivePrefix = "arXiv",
    primaryClass = "hep-ph",
    reportNumber = "PITT-PACC-2119",
    doi = "10.1103/PhysRevD.107.L091701",
    journal = "Phys. Rev. D",
    volume = "107",
    number = "9",
    pages = "L091701",
    year = "2023"
}

@article{Buchmuller:2003is,
    author = "Buchmuller, W. and Hamaguchi, Koichi and Ratz, Michael",
    title = "{Gauge couplings at high temperature and the relic gravitino abundance}",
    eprint = "hep-ph/0307181",
    archivePrefix = "arXiv",
    reportNumber = "DESY-03-078",
    doi = "10.1016/j.physletb.2003.09.017",
    journal = "Phys. Lett. B",
    volume = "574",
    pages = "156--161",
    year = "2003"
}

@article{Batell:2022qvr,
    author = "Batell, Brian and Ghalsasi, Akshay and Rai, Mudit",
    title = "{Dynamics of dark matter misalignment through the Higgs portal}",
    eprint = "2211.09132",
    archivePrefix = "arXiv",
    primaryClass = "hep-ph",
    reportNumber = "PITT-PACC-2213",
    doi = "10.1007/JHEP01(2024)038",
    journal = "JHEP",
    volume = "01",
    pages = "038",
    year = "2024"
}

@article{Chun:2021uwr,
    author = "Chun, Eung Jin",
    title = "{Bosonic dark matter in a coherent state driven by thermal fermions}",
    eprint = "2109.07423",
    archivePrefix = "arXiv",
    primaryClass = "hep-ph",
    doi = "10.1016/j.physletb.2022.136880",
    journal = "Phys. Lett. B",
    volume = "825",
    pages = "136880",
    year = "2022"
}

@article{Cyncynates:2024ufu,
    author = "Cyncynates, David and Simon, Olivier",
    title = "{Scalar Relics from the Hot Big Bang}",
    eprint = "2410.22409",
    archivePrefix = "arXiv",
    primaryClass = "hep-ph",
    doi = "10.1103/wjt8-9dh9",
    journal = "Phys. Rev. Lett.",
    volume = "135",
    number = "10",
    pages = "101003",
    year = "2025"
}

@article{Alachkar:2024crj,
    author = "Alachkar, Ahmad and Fairbairn, Malcolm and Marsh, David J. E.",
    title = "{Dilatonic Couplings and the Relic Abundance of Ultralight Dark Matter}",
    eprint = "2406.06395",
    archivePrefix = "arXiv",
    primaryClass = "hep-ph",
    doi = "10.1103/PhysRevLett.134.191003",
    journal = "Phys. Rev. Lett.",
    volume = "134",
    number = "19",
    pages = "191003",
    year = "2025"
}

@article{Cyncynates:2024bxw,
    author = "Cyncynates, David and Simon, Olivier",
    title = "{Minimal targets for dilaton direct detection}",
    eprint = "2408.16816",
    archivePrefix = "arXiv",
    primaryClass = "hep-ph",
    doi = "10.1103/8mc9-6cmr",
    journal = "Phys. Rev. D",
    volume = "112",
    number = "5",
    pages = "055002",
    year = "2025"
}

@article{Fermi-LAT:2015kyq,
    author = "Ackermann, M. and others",
    collaboration = "Fermi-LAT",
    title = "{Updated search for spectral lines from Galactic dark matter interactions with pass 8 data from the Fermi Large Area Telescope}",
    eprint = "1506.00013",
    archivePrefix = "arXiv",
    primaryClass = "astro-ph.HE",
    reportNumber = "FERMILAB-PUB-15-673-AE",
    doi = "10.1103/PhysRevD.91.122002",
    journal = "Phys. Rev. D",
    volume = "91",
    number = "12",
    pages = "122002",
    year = "2015"
}

@article{ODonnell:2024aaw,
    author = "O'Donnell, Kayla E. and Slatyer, Tracy R.",
    title = "{Constraints on dark matter with future MeV gamma-ray telescopes}",
    eprint = "2411.00087",
    archivePrefix = "arXiv",
    primaryClass = "hep-ph",
    reportNumber = "MIT-CTP/5792",
    doi = "10.1103/PhysRevD.111.083037",
    journal = "Phys. Rev. D",
    volume = "111",
    number = "8",
    pages = "083037",
    year = "2025"
}

@article{Essig:2013goa,
    author = "Essig, Rouven and Kuflik, Eric and McDermott, Samuel D. and Volansky, Tomer and Zurek, Kathryn M.",
    title = "{Constraining Light Dark Matter with Diffuse X-Ray and Gamma-Ray Observations}",
    eprint = "1309.4091",
    archivePrefix = "arXiv",
    primaryClass = "hep-ph",
    reportNumber = "YITP-SB-29-13, FERMILAB-PUB-13-377-A-T, MCTP-13-27",
    doi = "10.1007/JHEP11(2013)193",
    journal = "JHEP",
    volume = "11",
    pages = "193",
    year = "2013"
}

@article{EscuderoAbenza:2025cfj,
    author = "Escudero Abenza, Miguel and Hambye, Thomas",
    title = "{The simplest dark matter model at the edge of perturbativity}",
    eprint = "2505.02408",
    archivePrefix = "arXiv",
    primaryClass = "hep-ph",
    reportNumber = "CERN-TH-2025-087, ULB-TH/25-04",
    doi = "10.1016/j.physletb.2025.139696",
    journal = "Phys. Lett. B",
    volume = "868",
    pages = "139696",
    year = "2025"
}

@article{ParticleDataGroup:2024cfk,
    author = "Navas, S. and others",
    collaboration = "Particle Data Group",
    title = "{Review of particle physics}",
    doi = "10.1103/PhysRevD.110.030001",
    journal = "Phys. Rev. D",
    volume = "110",
    number = "3",
    pages = "030001",
    year = "2024"
}

@article{Silveira:1985rk,
    author = "Silveira, Vanda and Zee, A.",
    title = "{SCALAR PHANTOMS}",
    reportNumber = "DOE-ER-40048-13 P5",
    doi = "10.1016/0370-2693(85)90624-0",
    journal = "Phys. Lett. B",
    volume = "161",
    pages = "136--140",
    year = "1985"
}

@article{McDonald:1993ex,
    author = "McDonald, John",
    title = "{Gauge singlet scalars as cold dark matter}",
    eprint = "hep-ph/0702143",
    archivePrefix = "arXiv",
    reportNumber = "IFM-13-93",
    doi = "10.1103/PhysRevD.50.3637",
    journal = "Phys. Rev. D",
    volume = "50",
    pages = "3637--3649",
    year = "1994"
}

@article{Burgess:2000yq,
    author = "Burgess, C. P. and Pospelov, Maxim and ter Veldhuis, Tonnis",
    title = "{The Minimal model of nonbaryonic dark matter: A Singlet scalar}",
    eprint = "hep-ph/0011335",
    archivePrefix = "arXiv",
    reportNumber = "TPI-MINN-00-46, UMN-TH-1922-00, MCGILL-00-31, IASSNS-HEP-00-83",
    doi = "10.1016/S0550-3213(01)00513-2",
    journal = "Nucl. Phys. B",
    volume = "619",
    pages = "709--728",
    year = "2001"
}

@article{Batell:2026sml,
    author = "Batell, Brian and Ghalsasi, Akshay and Ghosh, Subhajit and Rai, Mudit",
    title = "{Phasing out Dark Matter Isocurvature with Thermal Misalignment}",
    eprint = "2603.18132",
    archivePrefix = "arXiv",
    primaryClass = "hep-ph",
    reportNumber = "PITT-PACC-2605, UT-WI-09-2026, MI-HET-881",
    month = "3",
    year = "2026"
}

@article{Gynther:2006wq,
    author = "Gynther, A.",
    title = "{Thermodynamics of electroweak matter}",
    eprint = "hep-ph/0609226",
    archivePrefix = "arXiv",
    reportNumber = "HU-P-D130",
    type = "Other thesis",
    month = "9",
    year = "2006"
}

@article{Gynther:2005dj,
    author = "Gynther, A. and Vepsalainen, M.",
    title = "{Pressure of the standard model at high temperatures}",
    eprint = "hep-ph/0510375",
    archivePrefix = "arXiv",
    reportNumber = "HIP-2005-46-TH",
    doi = "10.1088/1126-6708/2006/01/060",
    journal = "JHEP",
    volume = "01",
    pages = "060",
    year = "2006"
}

@article{Caputo:2022dkz,
    author = "Caputo, Andrea and Negro, Michela and Regis, Marco and Taoso, Marco",
    title = "{Dark matter prospects with COSI: ALPs, PBHs and sub-GeV dark matter}",
    eprint = "2210.09310",
    archivePrefix = "arXiv",
    primaryClass = "hep-ph",
    doi = "10.1088/1475-7516/2023/02/006",
    journal = "JCAP",
    volume = "02",
    pages = "006",
    year = "2023"
}

@article{Tomsick:2023aue,
    author = "Tomsick, John A. and others",
    title = "{The Compton Spectrometer and Imager}",
    eprint = "2308.12362",
    archivePrefix = "arXiv",
    primaryClass = "astro-ph.HE",
    doi = "10.22323/1.444.0745",
    journal = "PoS",
    volume = "ICRC2023",
    pages = "745",
    year = "2023"
}

@article{Orlando:2021get,
    author = "Orlando, Elena and others",
    title = "{Exploring the MeV sky with a combined coded mask and Compton telescope: the Galactic Explorer with a Coded aperture mask Compton telescope (GECCO)}",
    eprint = "2112.07190",
    archivePrefix = "arXiv",
    primaryClass = "astro-ph.HE",
    doi = "10.1088/1475-7516/2022/07/036",
    journal = "JCAP",
    volume = "07",
    number = "07",
    pages = "036",
    year = "2022"
}

@article{e-ASTROGAM:2017pxr,
    author = "Tavani, M. and others",
    editor = "De Angelis, A. and Tatischeff, V. and Grenier, I. A. and McEnery, J. and Mallamaci, M.",
    collaboration = "e-ASTROGAM",
    title = "{Science with e-ASTROGAM: A space mission for MeV{\textendash}GeV gamma-ray astrophysics}",
    eprint = "1711.01265",
    archivePrefix = "arXiv",
    primaryClass = "astro-ph.HE",
    doi = "10.1016/j.jheap.2018.07.001",
    journal = "JHEAp",
    volume = "19",
    pages = "1--106",
    year = "2018"
}

@article{Kierans:2020otl,
    author = "Kierans, Carolyn A.",
    collaboration = "AMEGO Team",
    title = "{AMEGO: Exploring the Extreme Multimessenger Universe}",
    eprint = "2101.03105",
    archivePrefix = "arXiv",
    primaryClass = "astro-ph.IM",
    doi = "10.1117/12.2562352",
    journal = "Proc. SPIE Int. Soc. Opt. Eng.",
    volume = "11444",
    pages = "1144431",
    year = "2020"
}

@article{Caputo:2022xpx,
    author = "Caputo, Regina and others",
    title = "{All-sky Medium Energy Gamma-ray Observatory eXplorer mission concept}",
    eprint = "2208.04990",
    archivePrefix = "arXiv",
    primaryClass = "astro-ph.IM",
    doi = "10.1117/1.JATIS.8.4.044003",
    journal = "J. Astron. Telesc. Instrum. Syst.",
    volume = "8",
    number = "4",
    pages = "044003",
    year = "2022"
}

@article{Dzhatdoev:2019kay,
    author = "Dzhatdoev, Timur and Podlesnyi, Egor",
    title = "{Massive Argon Space Telescope (MAST): A concept of heavy time projection chamber for $\gamma$-ray astronomy in the 100 MeV{\textendash}1 TeV energy range}",
    eprint = "1902.01491",
    archivePrefix = "arXiv",
    primaryClass = "astro-ph.HE",
    doi = "10.1016/j.astropartphys.2019.04.004",
    journal = "Astropart. Phys.",
    volume = "112",
    pages = "1--7",
    year = "2019"
}

@article{Hunter:2013wla,
    author = "Hunter, Stanley D. and others",
    title = "{A Pair Production Telescope for Medium-Energy Gamma-Ray Polarimetry}",
    eprint = "1311.2059",
    archivePrefix = "arXiv",
    primaryClass = "astro-ph.IM",
    doi = "10.1016/j.astropartphys.2014.04.002",
    journal = "Astropart. Phys.",
    volume = "59",
    pages = "18--28",
    year = "2014"
}

@article{Wu:2014tya,
    author = "Wu, Xin and Su, Meng and Bravar, Alessandro and Chang, Jin and Fan, Yizhong and Pohl, Martin and Walter, Roland",
    editor = "Takahashi, Tadayuki and den Herder, Jan-Willem A. and Bautz, Mark",
    title = "{PANGU: A High Resolution Gamma-ray Space Telescope}",
    eprint = "1407.0710",
    archivePrefix = "arXiv",
    primaryClass = "astro-ph.IM",
    doi = "10.1117/12.2057251",
    journal = "Proc. SPIE Int. Soc. Opt. Eng.",
    volume = "9144",
    pages = "91440F",
    year = "2014"
}

@article{GRAMS:2021tax,
    author = "Aramaki, Tsuguo and others",
    collaboration = "GRAMS",
    title = "{Overview of the GRAMS (Gamma-Ray AntiMatter Survey) Project}",
    doi = "10.22323/1.395.0653",
    journal = "PoS",
    volume = "ICRC2021",
    pages = "653",
    year = "2021"
}

@article{Fischer:2022pse,
    author = "Fischer, S. and Malyshev, D. and Ducci, L. and Santangelo, A.",
    title = "{New constraints on decaying dark matter from INTEGRAL/SPI}",
    eprint = "2211.06200",
    archivePrefix = "arXiv",
    primaryClass = "astro-ph.HE",
    doi = "10.1093/mnras/stad304",
    journal = "Mon. Not. Roy. Astron. Soc.",
    volume = "520",
    number = "4",
    pages = "6322--6334",
    year = "2023"
}

@article{Roach:2022lgo,
    author = "Roach, Brandon M. and Rossland, Steven and Ng, Kenny C. Y. and Perez, Kerstin and Beacom, John F. and Grefenstette, Brian W. and Horiuchi, Shunsaku and Krivonos, Roman and Wik, Daniel R.",
    title = "{Long-exposure NuSTAR constraints on decaying dark matter in the Galactic halo}",
    eprint = "2207.04572",
    archivePrefix = "arXiv",
    primaryClass = "astro-ph.HE",
    doi = "10.1103/PhysRevD.107.023009",
    journal = "Phys. Rev. D",
    volume = "107",
    number = "2",
    pages = "023009",
    year = "2023"
}

\end{document}